\documentclass[useAMS,usenatbib]{mn2e}

\usepackage{natbib}
\usepackage{graphicx}
\usepackage{multicol}
\usepackage{longtable}
\usepackage{amssymb}
\usepackage[width=1.0\textwidth]{caption}
\usepackage[usenames,dvipsnames]{color}
\RequirePackage[pdftex, plainpages = false, pdfpagelabels,pagebackref]{hyperref}
\RequirePackage{subfigure} %
\definecolor{Red}{rgb}{1.0,0,0}

\newcommand{\etal}    {{\it et al}}

\newcommand{\II}      {~{\sc ii}}
\newcommand{\III}     {~{\sc iii}}
\newcommand{\MB}      {Maxwell-Boltzmann}
\newcommand{\hf}      {Hf~2-2}

\title[The Continuum Emission Spectrum of \hf]{The Continuum Emission Spectrum of \hf\ near the Balmer Limit and the ORL versus CEL
abundance and temperature Discrepancy}
\author[P.J. Storey \& Taha
  Sochi]{P.J. Storey$^{1}$\thanks{E-mail:pjs@star.ucl.ac.uk (PJS).},
  Taha Sochi$^{1}$\footnotemark[1]\thanks{E-mail: t.sochi@ucl.ac.uk
    (TMS). Corresponding author.} \\
$^{1}$University College London,
  Department of Physics and Astronomy, Gower Street, London, WC1E 6BT}

\begin{document}

\date{Accepted XXX. Received XXX; in original form XXX}

\maketitle

\label{firstpage}

\begin{abstract}

\noindent The continuum spectrum of the planetary nebula Hf~2-2 close to the Balmer discontinuity
is modeled in the context of the long standing problem of the abundance and temperature discrepancy
found when analyzing optical recombination lines and collisionally excited forbidden lines in
nebulae. Models are constructed using single and double Maxwell-Boltzmann distributions as well as
$\kappa$-distributions for the energies of the free electrons. New results for the necessary
continuum and line emission coefficients are presented calculated with $\kappa$-distributed
energies. The best fit to the observed continuum spectrum is found to be a model comprising two
components with dramatically different temperatures and with a Maxwell-Boltzmann distribution of
electron energies. On the basis of a $\chi^2$ analysis, this model is strongly favored over a model
with $\kappa$-distributed electron energies.
\end{abstract}

\begin{keywords}
atomic transition -- atomic spectroscopy -- Hf~2-2 -- planetary nebulae -- nebular physics --
optical recombination lines -- collisionally excited lines -- Balmer continuum -- abundance
discrepancy -- electron temperature -- electron distribution -- Maxwell-Boltzmann -- kappa
distribution.

\vspace{0.5cm}Note: figures are colored in the online version.
\end{keywords}

\section{Introduction} \label{Introduction}

The long standing problem in nebular physics related to the abundance and temperature discrepancy
between the results obtained from optical recombination lines (ORL) and collisionally excited lines
(CEL) has been visited by many researchers in the last few decades. Several explanations for the
general trend of obtaining higher abundances and lower temperatures from ORL than from CEL of the
same ions have been put forward. One explanation is a multi-component model of planetary nebulae in
which low temperature, high metallicity components (clumps) produce the ORL and are embedded in a
high temperature, relatively low metallicity, hydrogen-rich component that produces the CEL
\citep{LiuSBDCB2000}. It has been suggested recently \citep{NichollsDS2012} that this discrepancy
may be largely explained by the assumption of a different electron energy distribution, a
$\kappa$-distribution, rather than the \MB\ (MB) distribution which is traditionally accepted as
the dominant distribution in the low density plasma found in the planetary nebulae.

The debate about the electron energy distribution in astronomical objects, including planetary
nebulae, is relatively old and dates back to at least the 1940s when Hagihara \cite{Hagihara1944}
suggested that the distribution of free electrons in gaseous assemblies deviates considerably from
the MB distribution. This was refuted by \citet{BohmA1947} who, on the basis of a detailed
quantitative balance analysis, concluded that any deviation from the Maxwellian equilibrium
distribution is very small. The essence of their argument is that for typical planetary nebulae
conditions, the thermalization process of elastic collisions is by far the most frequent event and
typically occurs once every second, while other processes that shift the system from its
thermodynamic equilibrium, like inelastic scattering with other ions that leads to metastable
excitation or recapture, occur at much larger time scales estimated to be months or even years.
Bohm and Aller also indicated the significance of any possible deviation from a Maxwellian
distribution on derived elemental abundances. However, nobody seems to have considered non-MB
electron distributions as a possible solution for the ORL/CEL discrepancy problem until the recent
proposal of \citet{NichollsDS2012}. In the light of this suggestion, it seems appropriate to ask
whether there is any empirical evidence from observation for departures from MB energy
distributions in nebulae.

A preliminary investigation of this possibility has been given recently by \citet{SochiThesis2012} and
\citet{StoreySTemp2013} who proposed a method based on using atomic dielectronic recombination (DR)
theoretical data in conjunction with observational line fluxes to sample the free electron
distribution and compare to the theoretical distributions, i.e. \MB\ and $\kappa$. For all the
objects studied by \citet{StoreySTemp2013} the limited observational data analyzed were found to be
better described by a single MB distribution than a $\kappa$-distribution but the uncertainties
were sufficient that a $\kappa$-distribution could not be conclusively excluded. A two-component
model incorporating low temperature material was also found to not fit the data as well as a single
MB model.

The current paper approaches this problem by modeling the magnitude of the Balmer discontinuity and
nearby Balmer continuum of the planetary nebula \hf. This nebula is an extreme example of the
ORL/CEL discrepancy trend since its abundance discrepancy factor (ADF), defined as the ratio of ORL
abundance to CEL abundance, reaches an exceptionally high value between 68-84
\citep{Liu2003,WessonLB2003,ZhangLWSLD2004,LiuLBL2004,WessonL2004,LiuBZBS2006,WessonBLSED2008}. An
exceptionally high temperature discrepancy between the CEL and ORL results, which differ by about
an order of magnitude, has also been obtained for this nebula.

The shape of the Balmer continuum is principally determined by a convolution of the $n=2$ hydrogen
recombination cross-section and the free electron energy distribution. The shape of the continuum
therefore offers the possibility of gaining information about the free electron energy
distribution. We shall see below that in the case of \hf\ the continuum shape is primarily
determined by the free electron energy distribution. In section~\ref{Hf2-2} we outline the
properties of \hf\ and in section~\ref{Theory} we deal with the theory describing the continuum
processes and new results relating to recombination using a $\kappa$-distribution. The fitting
procedure and the resulting fits to the spectral data with MB and $\kappa$ distributions are
described in section~\ref{Method} and our conclusions in section~\ref{Conclusions}.

\section{\hf} \label{Hf2-2}

\hf\ is a faint southern highly symmetric low density planetary nebula with a central cavity and
interior disc-shaped multi-shell structure. Its spectrum includes carbon, nitrogen, oxygen, neon,
sulphur and argon lines as well as helium. \hf\ seems to have a high abundance in He and C with a
high C/O ratio and an exceptionally strong C\II\ $\lambda$4267~\AA\ signature. The nebula, which is
in the galactic bulge at a distance of about 4.0-4.5~kpc, has a spectrally-varying close binary
central system with a period of about 0.40~day \citep{Bond2000}. \hf\ has common spectral features
with old novae like DQ~Her \citep{CahnK1971,Maciel1984,Kaler1988,Liu2003,LiuBZBS2006,SchaubBH2012},
and therefore being a planetary nebula may be disputed.

This unusual nebula is marked by a number of exceptional features. \hf\ exhibits a very high ADF,
possibly the highest recorded for a planetary nebula, with very strong ORL emission. \hf\ also has
the very large C/O abundance ratio of about 19 \citep{PatriarchiP1994}. A third unusual feature is
a very large Balmer jump associated with a rapid fall in the Balmer continuum intensity towards
shorter wavelengths. An estimation based on the magnitude of the Balmer discontinuity indicates an
electron temperature of about 780-1000~K in sharp contrast to the forbidden lines estimation of
about 8800~K from an [O\III] line
\citep{Liu2003,LiuLBL2004,ZhangLWSLD2004,LiuBZBS2006,McnabbFLBS2012}. The deduced Balmer
discontinuity temperature is one of the lowest observed for a planetary nebula. These findings led
\citet{LiuBZBS2006} to conclude that a multi-component model of cold hydrogen-deficient knots
embedded into a metal-poor nebula may be inevitable to explain these exceptional observations.

With regard to ORL electron temperature derivations for \hf, \citet{BastinThesis2006} obtained a
mean electron temperature of $<600$~K from C\II\ lines, while \citet{Liu2006} derived an electron
temperature of about 630~K and \citet{McnabbFLBS2012} a temperature of about 3160~K from O\II\
lines. As for the stellar temperature of this nebula, it is estimated by \citet{LiuBZBS2006}
between 50000--67000~K, while \citet{SchaubBH2012} set the temperature of the two stars (assuming a
binary system which seems to be largely accepted) in their model to 67000~K and 7500~K.

\begin{figure}
\centering{}
\includegraphics [scale=0.7] {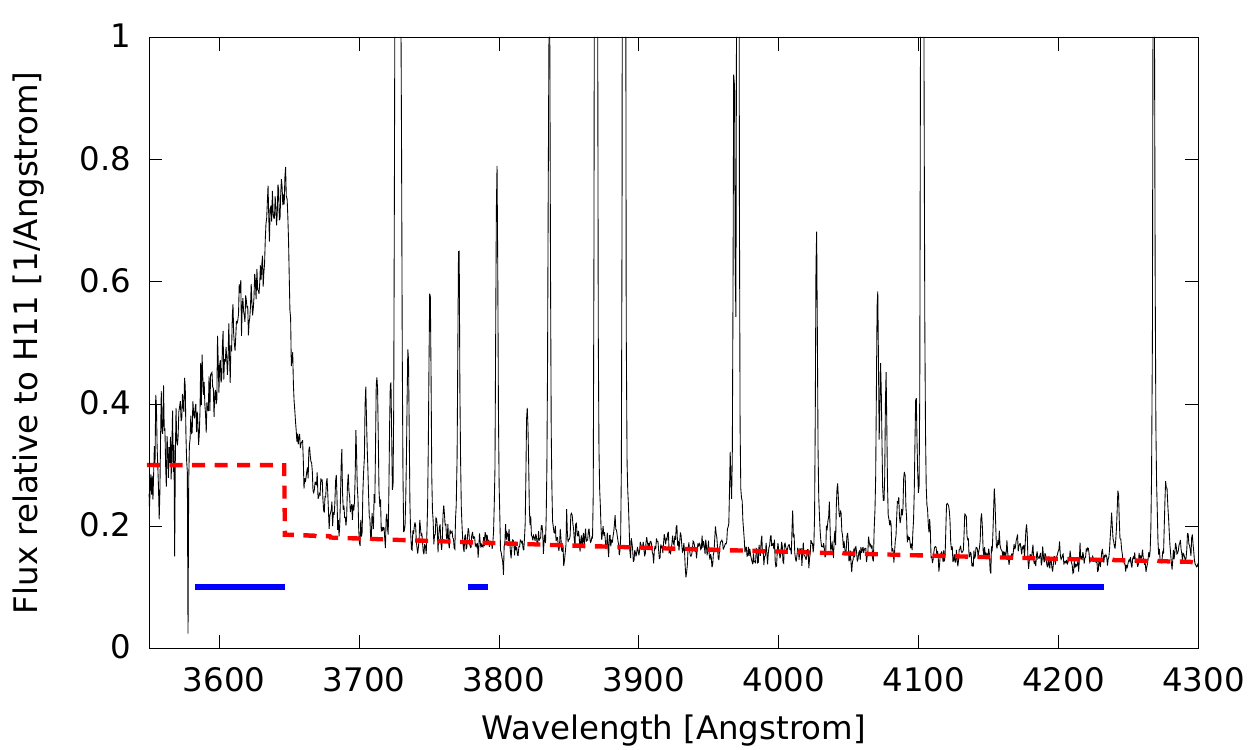} \caption{
The observed \hf\ spectrum (solid) and the continuum fit with a single \MB\ distribution with
$T=8800$~K (dashed). The fit was optimized on the two longer wavelength of the three wavelength
segments (shown in the figure as horizontal lines): 3585-3645, 3780-3790 and 4180-4230~\AA.}
\label{ContFitSeg}
\end{figure}

The observational data of \hf\ which is used in the present paper were obtained in 2001 by a Boller
and Chivens long-slit spectrograph mounted on the ground-based 1.52~m European Southern Observatory
telescope located in La Silla Chile. More details about this data set can be found in
\citet{LiuBZBS2006}. The observed spectrum is shown in Figure~\ref{ContFitSeg}, with the measured
flux normalized to Balmer H11, an apparently unblended line close to the Balmer discontinuity.

\section{Theory} \label{Theory}

To model the magnitude of the Balmer discontinuity and the shape of the Balmer continuum we
consider the contributions to the continuum spectrum from recombination of H$^+$ and He$^+$, H
2-photon emission and the underlying scattered stellar continuum. Recombination of He$^{2+}$ is
neglected due to the very low He$^{2+}$/H$^+$ fraction in \hf\ which is estimated to be about 0.002
\citep{ZhangLWSLD2004,LiuBZBS2006}.

The recombination of an atomic ion $X^+$ with an electron, $e^-$ of energy $E$ to a state $X^*$ of the recombined ion,
\begin{equation}
 X^+ + e^- \rightarrow X^* +h\nu,\label{genrecomb}
\end{equation}
gives rise to continuous emission with emission coefficient (energy emitted per unit time per unit
frequency and per unit particle densities)
\begin{equation}
\gamma^*(\nu) = \frac{h^3 \alpha^3 c}{32\pi^2 m_e a_0^2} \frac{\omega^*}{\omega^+} \left(
\frac{h\nu}{R} \right)^3 \left( \frac{R}{E} \right)^{\frac{1}{2}} f(E)\ \sigma^*_{\nu} \label{gamma}
\end{equation}
where, $\nu$ is the frequency of the emitted photon, $\omega^+$ and $\omega^*$ are the statistical
weights of the initial and final states respectively, $f(E)$ is the free electron energy
distribution function, $\sigma^*_{\nu}$ is the photoionization cross-section for the inverse
process of Equation~\ref{genrecomb} and  $R$ is the Rydberg energy constant. The other symbols have
their usual meanings. In the present paper, the photoionization cross-sections for states of H and
He are taken from \citet{StoreyH91} and \citet{HummerS98} respectively as described in more detail
by \citet{ErcolanoS06}.

In this paper we consider two possible forms for $f(E)$, the Maxwell-Boltzmann distribution
\begin{equation}
f_{_{\rm MB}}(E,T) =
\frac{2}{\left(kT\right)^{3/2}}\sqrt{\frac{E}{\pi}}e^{-\frac{E}{kT}}\label{ElDiEq2},
\end{equation}
where $T$ is the electron temperature, and the $\kappa$-distribution \citep{Vasyliunas1968, SummersT91} given
by

{\scriptsize
\begin{equation}
f_{\kappa}\left(E,
T_{\kappa}\right)=\frac{2\sqrt{E}}{\sqrt{\pi(kT_{\kappa})^3}}\frac{\Gamma\left(\kappa+1\right)}{(\kappa-\frac{3}{2})^{\frac{3}{2}}\Gamma\left(\kappa-\frac{1}{2}\right)}\left(1+\frac{E}{(\kappa-\frac{3}{2})
kT_{\kappa}}\right)^{-\left(\kappa+1\right)} \label{kappaEq}
\end{equation}}
\noindent %
where $\kappa$ is a parameter defining the distribution, $\Gamma$ is the gamma-function for the
given arguments, and $T_{\kappa}$ is a characteristic temperature. Note that $f_{\kappa}$ becomes a
MB distribution with $T_{\kappa} \rightarrow T$ as $\kappa \rightarrow \infty$.

The normalization of the continuum flux to H11 flux requires that the effective recombination
coefficients for H11 should be calculated both with a Maxwell-Boltzmann distribution and a
$\kappa$-distribution. The hydrogen line emissivities tabulated by \citet{HummerS87} and
\citet{StoreyH95} were calculated assuming that the free electron energy distribution is described
by a Maxwell-Boltzmann distribution for all physical processes between bound and continuum states
involving free electrons. Here we use the techniques and computer codes described in the last two
references and extend them to include a $\kappa$-distribution. At the electron number densities
typical of photoionized nebulae ($10^2-10^5$~cm$^{-3}$) the populations of the low-lying states of
H are determined primarily by recombination and radiative cascading and are relatively insensitive
to the ambient electron density. Hence any error introduced into the calculation of collision rates
between high-$n$ states caused by the use of a Maxwell-Boltzmann distribution rather than a
$\kappa$-distribution should have minimal effect at nebular densities. We therefore make the
approximation of computing the direct recombination coefficients to all the individual levels using
a $\kappa$-distribution but retain a Maxwell-Boltzmann distribution for the energy and angular
momentum changing collisions among the higher-$n$ states. This should provide a good approximation
for the H11 emissivity.

For a general free electron energy distribution, the recombination coefficient to a state X$^*$ is
given by

\begin{equation}
\alpha^*_{RC} =  \frac{R^{\frac{5}{2}}}{\sqrt{2}c^2 m_e^{\frac{3}{2}}} \frac{\omega^*}{\omega^+}
\int_0^\infty \left( \frac{h\nu}{R} \right)^2 \left( \frac{R}{E} \right)^{\frac{1}{2}}
\sigma^*_{\nu}\ f(E)\ {\rm d}\left(\frac{E}{R} \right) \label{alphaEq}
\end{equation}
On solving the collisional-radiative recombination problem for hydrogen we obtain the recombination
coefficients to all levels and the effective recombination coefficients $\alpha_{\rm eff}(\lambda)$
for a transition of wavelength $\lambda$. We define a line emission coefficient $\epsilon(\lambda)$
as the energy emitted per unit volume per unit time for unit ion and electron density, so that
\begin{equation}
\epsilon(\lambda) = \alpha_{\rm eff}(\lambda)\ \frac{hc}{\lambda}.
\end{equation}

\citet{NichollsDSKP2013} derive the following approximate expression for converting recombination
coefficients calculated with a Maxwell-Boltzmann electron energy distribution to those applicable
with a $\kappa$-distribution,
\begin{equation}
 x(\kappa) \equiv \frac{\alpha_{\kappa}(\lambda)}{\alpha_{\rm MB}(\lambda)} =  \frac{\epsilon_{\kappa}(\lambda)}{\epsilon_{\rm MB}(\lambda)}
=  \frac{(1-\frac{3}{2\kappa})\Gamma(\kappa+1)}{(\kappa-\frac{3}{2})^{\frac{3}{2}}\Gamma(\kappa-\frac{1}{2})}.\label{MBkappa}
\end{equation}
To obtain this result, \citet{NichollsDSKP2013} use the fact that the photoionization cross-section
falls approximately as $(\nu_0/\nu)^3$ for $\nu > \nu_0$, where $\nu_0$ is the threshold frequency
for photoionization. so that the integrand in Equation~\ref{alphaEq} contains a $1/\nu$ term.
\citet{NichollsDSKP2013} simplify the integral by moving the $1/\nu$ term outside of the integral,
making it possible to carry out the integration analytically. In practice the integral is a
convolution of the free-electron energy distribution with the $1/\nu$ term, which is energy
dependent. At low temperatures the rapid fall in the electron energy distribution function with
increasing energy means that it is a good approximation to neglect the energy dependence of the
frequency term. At higher temperatures, both terms are essential in the calculation of the
recombination coefficient. The expression for $x(\kappa)$ in Equation~\ref{MBkappa} is independent
of temperature, density and transition. This approximation will fail at sufficiently high
temperatures where the effects of the free-electron energy distribution and the frequency dependent
term become comparable.

Using the approximate function $x(\kappa)$, we can express the results of our more complete
collisional-radiative treatment for $\epsilon(\lambda)$ in terms of a correction factor
$y(\lambda,T,\kappa)$ as follows

\begin{equation}
\epsilon_{\kappa}(\lambda) = \epsilon_{\rm MB}(\lambda)\ x(\kappa)\ y(\lambda,T,\kappa).
\label{epsEq}
\end{equation}
The values of $\epsilon_{\rm MB}$ can be obtained from \citet{HummerS87} and \citet{StoreyH95}. In
Table~\ref{H11MBkappa} we tabulate values of $y(\lambda,T,\kappa)$ for H11 corresponding to various
values of $\kappa$ at a range of temperatures. We also tabulate in the last line $x$ as a function
of $\kappa$. Note that, in principle, $y$ also depends on electron density but in practice it is
very insensitive to electron density for densities up to $10^5$~cm$^{-3}$, so we only tabulate
values calculated at $10^3$~cm$^{-3}$.

\begin{table}
\caption{Values of the correction factor $y(\lambda,T,\kappa)$, defined in Equation~\ref{epsEq}, as
a function of $\kappa$ and logT[K] for computing the emission coefficient,
$\epsilon_{\kappa}(\lambda)$ for H11. In the last row of the table $x$, as defined in Equation
\ref{MBkappa}, is given as a function of $\kappa$. \label{H11MBkappa}}
\begin{center} \vspace{-0.3cm}
\begin{tabular}{cllllllll}
\hline
   logT[K] & \multicolumn{7}{c}{$\kappa$} \\
           & 50.0 & 20.0 & 10.0 & 7.0 & 5.0 & 3.0 & 2.0 \\
\hline
     2.6 &    0.994 &    0.994 &    0.996 &    0.997 &    0.999 &    1.004 &    1.017 \\
     2.7 &    0.994 &    0.995 &    0.997 &    0.999 &    1.002 &    1.011 &    1.031 \\
     2.8 &    0.995 &    0.996 &    0.999 &    1.001 &    1.005 &    1.017 &    1.044 \\
     2.9 &    0.995 &    0.997 &    1.001 &    1.004 &    1.009 &    1.024 &    1.057 \\
     3.0 &    0.996 &    0.998 &    1.002 &    1.007 &    1.013 &    1.031 &    1.071 \\
     3.1 &    0.996 &    0.999 &    1.004 &    1.009 &    1.016 &    1.038 &    1.084 \\
     3.2 &    0.996 &    1.000 &    1.005 &    1.011 &    1.020 &    1.045 &    1.099 \\
     3.3 &    0.996 &    1.000 &    1.007 &    1.013 &    1.023 &    1.053 &    1.115 \\
     3.4 &    0.996 &    1.000 &    1.008 &    1.016 &    1.027 &    1.062 &    1.133 \\
     3.5 &    0.995 &    1.001 &    1.010 &    1.019 &    1.032 &    1.072 &    1.154 \\
     3.6 &    0.995 &    1.001 &    1.012 &    1.022 &    1.037 &    1.083 &    1.177 \\
     3.7 &    0.995 &    1.001 &    1.014 &    1.026 &    1.043 &    1.096 &    1.203 \\
     3.8 &    0.995 &    1.003 &    1.017 &    1.031 &    1.051 &    1.112 &    1.235 \\
     3.9 &    0.995 &    1.004 &    1.020 &    1.036 &    1.059 &    1.128 &    1.269 \\
     4.0 &    0.994 &    1.004 &    1.022 &    1.040 &    1.066 &    1.144 &    1.306 \\
     4.1 &    0.995 &    1.006 &    1.027 &    1.047 &    1.076 &    1.165 &    1.350 \\
     4.2 &    0.995 &    1.008 &    1.032 &    1.054 &    1.087 &    1.187 &    1.398 \\
     4.3 &    0.995 &    1.010 &    1.036 &    1.060 &    1.097 &    1.210 &    1.450 \\
\hline
  $x(\kappa)$     & 1.008 & 1.020 & 1.043 & 1.066 & 1.103 & 1.228 & 1.596 \\
\hline
\end{tabular}
\end{center}
\end{table}

As can be seen, the values from Equation~\ref{MBkappa} are a good approximation for large values of
$\kappa$ but the ratio increases as $\kappa$ decreases. In addition, the values from
Equation~\ref{MBkappa} are generally smaller than the more exact values, with the difference being
largest for the smallest tabulated $\kappa$ and the highest temperatures, as expected.

Since our model of the recombination process with $\kappa$-distributed electron energies does not
treat energy and angular momentum changing collisions correctly we cannot use the results to model
the high Balmer lines as was done for example by \citet{ZhangLWSLD2004} using the code originally
authored by one of us (PJS). We therefore restrict the current model to calculation of the
continuum processes only, normalized to the H11 flux.

Our model of the continuum also includes the hydrogen 2-photon emission. The emission coefficient
for the hydrogen 2s-1s 2-photon emission is

\begin{equation}
\epsilon_{2q}(\nu) = \alpha({\rm 2s}) \frac{A_{2q}(\nu)}{A_{2q}+C_{{\rm 2s,2p}}\ N_p}\ h\nu,
\label{2q}
\end{equation}
where $\alpha({\rm 2s})$ is the total recombination coefficient to the 2s state of H, $A_{2q}$ is
the total 2s-1s spontaneous transition probability, $A_{2q}(\nu)$ is the probability per unit
frequency, $C_{{\rm 2s,2p}}$ is the coefficient for proton collisional transitions between the 2s
and 2p states and $N_p$ is the proton number density. We assume that the population of the 2p state
is negligible at the relevant densities \citep{HummerS87}, and we take $C_{{\rm 2s,2p}}$ from
\citet{Seaton55} and $A_{2q}$ and $A_{2q}(\nu)$ from \citet{NussbaumerS84}. Values of $\alpha({\rm
2s})$ for an MB distribution were provided by \citet{HummerS87} and \citet{StoreyH95}. From our new
calculation of the hydrogen collisional-radiative problem with a $\kappa$-distribution we obtain
values of $\alpha({\rm 2s})$ as a function of $\kappa$ and temperature. These values can be related
to those obtained with an MB distribution  in the same way as in Equation~\ref{epsEq}, via the
function $x(\kappa)$
\begin{equation}
\alpha_{\kappa}({\rm 2s}) = \alpha_{\rm MB}({\rm 2s})\ x(\kappa)\ z({\rm 2s},T,\kappa) \label{zdef}
\end{equation}
where we tabulate the correction factors $z({\rm 2s},T,\kappa)$ in Table~\ref{2sMBkappa}.

\begin{table}
\caption{Values of the correction factor $z({\rm 2s},T,\kappa)$, defined in Equation~\ref{zdef}, as
a function of $\kappa$ and logT[K] for computing the total recombination coefficient to the 2s
state of hydrogen with $\kappa$-distributed electron energies.
  \label{2sMBkappa}}
\begin{center} \vspace{-0.3cm}
\begin{tabular}{cllllllll}
\hline
   logT[K] & \multicolumn{7}{c}{$\kappa$} \\
           &  50.0 & 20.0 & 10.0 & 7.0 & 5.0 & 3.0 & 2.0 \\
\hline
     2.6 &    0.994 &    0.994 &    0.995 &    0.995 &    0.996 &    1.000 &    1.009 \\
     2.7 &    0.994 &    0.994 &    0.996 &    0.997 &    0.999 &    1.005 &    1.020 \\
     2.8 &    0.994 &    0.995 &    0.997 &    0.999 &    1.001 &    1.009 &    1.029 \\
     2.9 &    0.995 &    0.996 &    0.998 &    1.000 &    1.004 &    1.014 &    1.038 \\
     3.0 &    0.995 &    0.996 &    0.999 &    1.002 &    1.006 &    1.018 &    1.047 \\
     3.1 &    0.995 &    0.996 &    1.000 &    1.003 &    1.007 &    1.022 &    1.054 \\
     3.2 &    0.994 &    0.996 &    1.000 &    1.003 &    1.009 &    1.025 &    1.062 \\
     3.3 &    0.994 &    0.996 &    1.001 &    1.004 &    1.011 &    1.029 &    1.070 \\
     3.4 &    0.994 &    0.996 &    1.001 &    1.005 &    1.013 &    1.033 &    1.079 \\
     3.5 &    0.994 &    0.997 &    1.002 &    1.007 &    1.015 &    1.038 &    1.089 \\
     3.6 &    0.994 &    0.998 &    1.004 &    1.009 &    1.018 &    1.045 &    1.101 \\
     3.7 &    0.995 &    0.999 &    1.006 &    1.012 &    1.022 &    1.052 &    1.114 \\
     3.8 &    0.998 &    1.001 &    1.010 &    1.017 &    1.028 &    1.062 &    1.131 \\
     3.9 &    0.997 &    1.002 &    1.011 &    1.019 &    1.032 &    1.069 &    1.147 \\
     4.0 &    0.994 &    1.000 &    1.010 &    1.019 &    1.033 &    1.075 &    1.162 \\
     4.1 &    0.994 &    1.000 &    1.011 &    1.022 &    1.038 &    1.085 &    1.182 \\
     4.2 &    0.995 &    1.002 &    1.014 &    1.026 &    1.044 &    1.096 &    1.207 \\
     4.3 &    0.994 &    1.001 &    1.015 &    1.028 &    1.048 &    1.107 &    1.230 \\
\hline
\end{tabular}
\end{center}
\end{table}

\section{Modeling the continuum} \label{Method}

The observed spectrum of \hf\ with intensity normalized relative to H11 is shown in
Figure~\ref{ContFitSeg}. The figure also shows the spectral segments that we use for fitting,
chosen to be as free as possible from significant spectral lines. The wavelength ranges for the
three segments are $\lambda=$ 3585-3645, 3780-3790 and 4180-4230~\AA.

\subsection{The continuum longward of the Balmer edge}

The observed continuum longward of the Balmer edge is comprised principally of hydrogen free-bound
emission with principal quantum number $n>2$, helium free-bound emission, hydrogen 2-photon
emission and scattered starlight from the central star of the nebula. The emissivity of the H
2-photon emission relative to H11 is relatively insensitive to the temperature of the emitting
material and comprises up to 10\% of the continuum in this spectral region. The contribution from
$n>2$ H free-bound emission varies strongly with the temperature, being 3\% just longward of the
Balmer edge at $T=10^4$~K and falling rapidly as the temperature decreases. We model the remaining
background contribution to the flux relative to H11 with an empirical power-law distribution

\begin{equation}
F(\lambda) = F_0\ \left( \frac{3647}{\lambda} \right)^{\beta} \label{power}
\end{equation}
where $F_0$ is the contribution at the Balmer edge and wavelengths in \AA\ are vacuum wavelengths.

The parameters $F_0$ and $\beta$ were obtained by fitting the model continuum to the two longer
wavelength segments shown in Figure~\ref{ContFitSeg} by minimizing the rms deviations of the fit
from the observed spectrum. In computing the model continuum we take $N({\rm He}^+)$/$N({\rm H}^+)$
= 0.103 \citep{LiuBZBS2006}. The same authors quote $N({\rm He}^{2+})$/$N({\rm H}^+)$ = 0.002,
which means that He$^{2+}$ recombination makes a negligible contribution to the continuum, so we
neglect this component. The derived continuum is very weakly sensitive to the electron number
density through the 2-photon component (Equation~\ref{2q}) and we adopt $N_e= 1000$~cm$^{-3}$.

The best fit values of $F_0$ and $\beta$ were derived for two models. The first uses a MB
distribution for the free electrons with a temperature of 8800~K chosen to reflect the forbidden
line values derived by \citet{LiuBZBS2006}. In any model comprising two MB distributions, the low
temperature component will contribute a negligible amount to the continuum longward of the Balmer
edge, so it is appropriate to use the forbidden line temperature to determine the continuum
component in this wavelength range.

As we shall see below, the best fit to the continuum using a single $\kappa$-distribution yields an
electron temperature of a few thousand Kelvin and a small value of $\kappa$. So for our second
model of the underlying continuum we adopt a $\kappa$-distribution with $\kappa=2$ and $T=3000$~K.
The results of the two underlying continuum models are summarized in Table~\ref{confit}. The values
of $F_0$ and $\beta$ are relatively insensitive to the assumptions about the electron energy
distribution and its temperature. To illustrate the fit longward of the Balmer edge, we show in
Figure~\ref{ContFitSeg} the full fit to the two longer wavelength segments of the continuum
including all the processes described above, using the parameters (a) from Table~\ref{confit} and
assuming a single MB distribution with $T=8800$~K.

\begin{table}
\caption{Underlying continuum fitting parameters $F_0$ and $\beta$ for two scenarios (a) A single
MB distribution with temperature 8800~K, and (b) a single $\kappa$-distribution with $\kappa=2$ and
$T =3000$~K. \label{confit}}
\begin{center} 
\begin{tabular}{|l|c|c|}
\hline
  &   $F_0$[\AA$^{-1}$]  & $\beta$ \\
\hline
  a  & 0.161 & 1.588 \\
  b  & 0.174 & 1.536 \\
\hline
\end{tabular}
\end{center}
\end{table}

\subsection{Fits with Maxwell-Boltzmann distributions}

\begin{figure}
\centering{}
\includegraphics [scale=0.7] {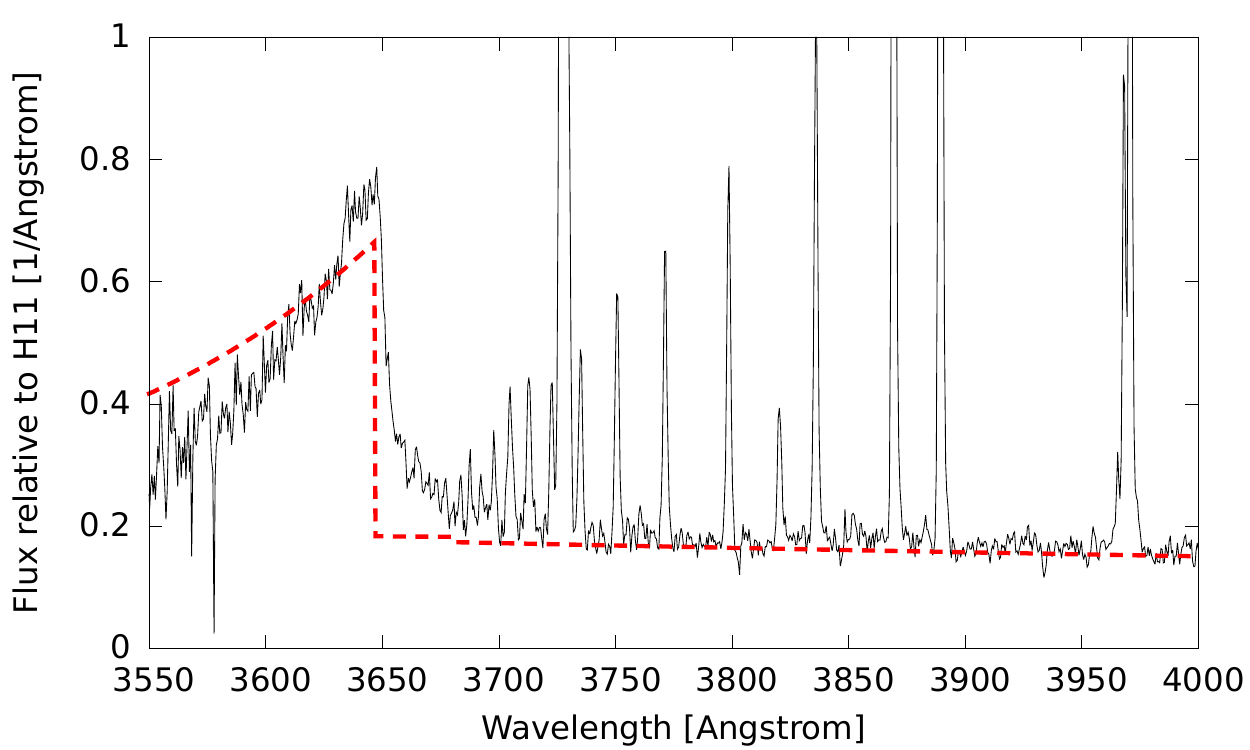}
\caption{The observed \hf\ spectrum (solid) and the single-component Maxwell-Boltzmann fit (dashed)
where the temperature is treated as a free parameter. The final optimized temperature is
$T=1334$~K.} \label{SinglefreeMB}
\end{figure}

It is clear from Figure~\ref{ContFitSeg} that the magnitude of the Balmer discontinuity and the
slope of the Balmer continuum shortward of the discontinuity cannot be explained by emission from
hydrogen and helium at the forbidden line temperature. This has already been discussed by
\citet{LiuBZBS2006} who estimated that the shape of the continuum implies that the emitting region
has a temperature of $\approx 900$~K. In Figure~\ref{SinglefreeMB} we show the result of making a
fit to all three spectral segments shown in Figure~\ref{ContFitSeg} using a single MB electron
energy distribution in which the temperature is a free parameter. The background continuum is
computed using the parameters from fit (a) in Table~\ref{confit}. The temperature of best fit is
1334~K and although the continuum model is improved compared to Figure~\ref{ContFitSeg}, it matches
neither the magnitude of the discontinuity nor the slope adequately.

\begin{figure}
\centering{}
\includegraphics [scale=0.7] {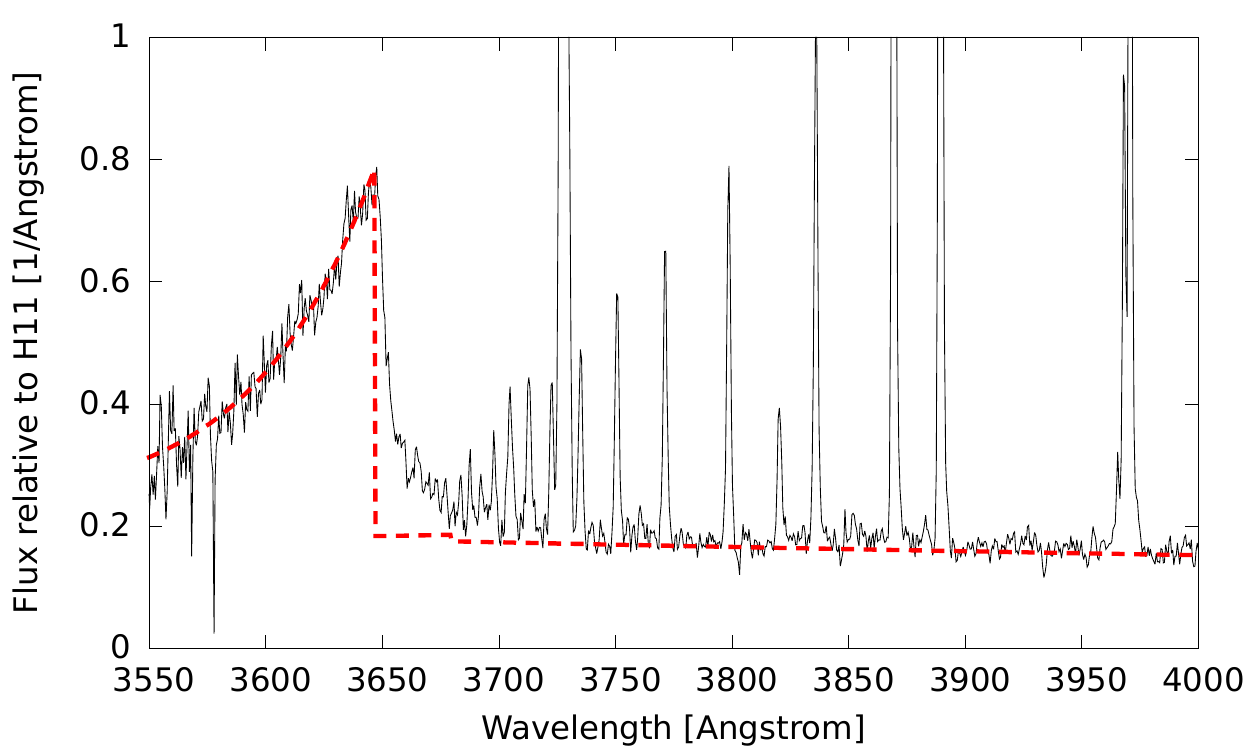}
\caption{The observed \hf\ spectrum (solid) and a two-component \MB\ fit (dashed) with two
temperatures: one is the forbidden line temperature which is fixed at $T=8800$~K and the other is
free to vary. The ratio of the volumes of the two components is also treated as a free parameter.}
\label{2MB1free}
\end{figure}

As discussed earlier, one resolution of the ORL/CEL abundance and temperature discrepancy in
nebulae that has been proposed is that there are relatively cold high-metallicity knots embedded in
hotter lower abundance gas. We therefore attempt to match the continuum with two components having
different temperatures. We set the temperature of one component equal to a typical forbidden line
temperature for this object of 8800~K and allow the other to vary. The continuum emissivity has
only weak density dependence so we choose representative electron densities for the two components
of 1000~cm$^{-3}$ for the forbidden line region and 5000~cm$^{-3}$ for the cold material \citep{LiuBZBS2006}.
The relative emission measure of the two components is then allowed to vary by making the volume
ratio of the two components a free parameter. In Figure~\ref{2MB1free} we show the resulting fit,
which is excellent. The optimum temperature of the cold component is 540~K and the fraction of the
total volume occupied by the cold component is 0.00706. Hence the cold component has an emission
measure, which is proportional to the product of the volume fraction and particle density squared,
that is 15.0\% of the total recombination line and continuum emission measure.

\subsection{Fit with $\kappa$-distribution}

\begin{figure}
\centering{}
\includegraphics
[scale=0.7] {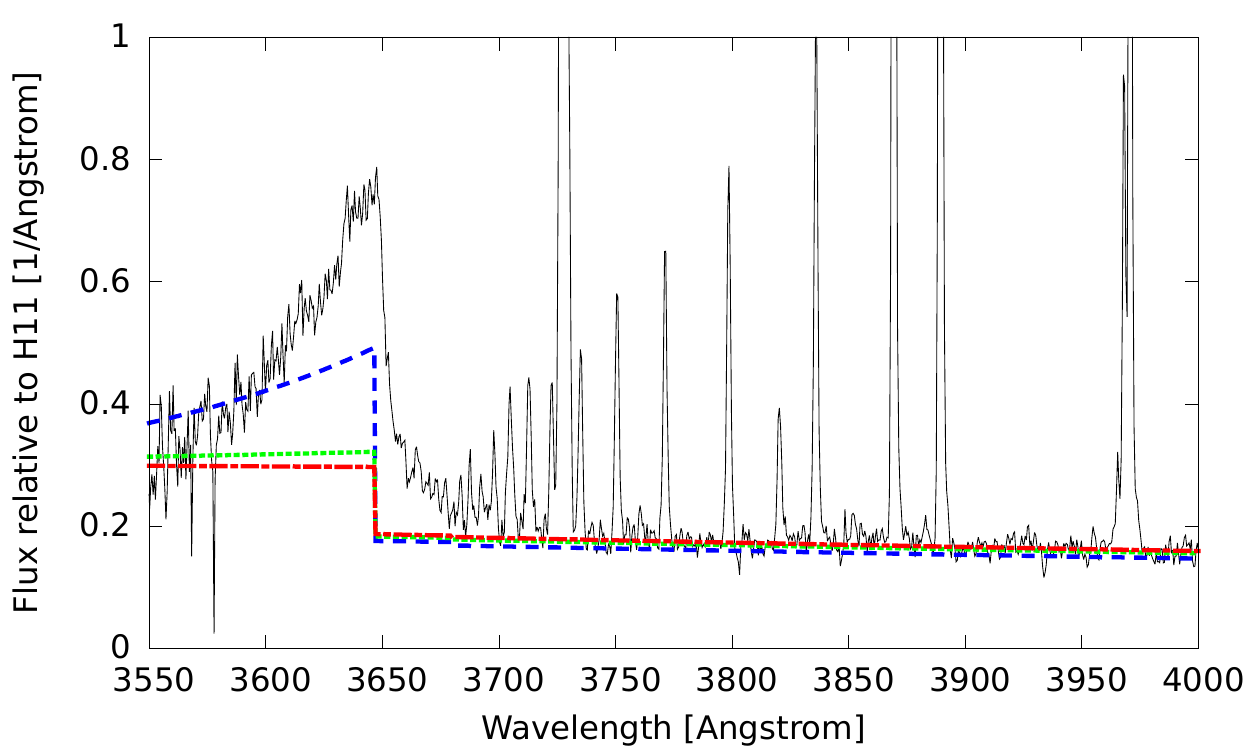}  \caption{Model continua with a $\kappa$-distribution of electron
energies for a temperature $T=10000$~K; $\kappa=2$ (dashed blue line), $\kappa=5$ (dotted green
line), and $\kappa=25$ (dot-dashed red line).} \label{T100003kappa}
\end{figure}

\begin{figure}
\centering{}
\includegraphics
[scale=0.7] {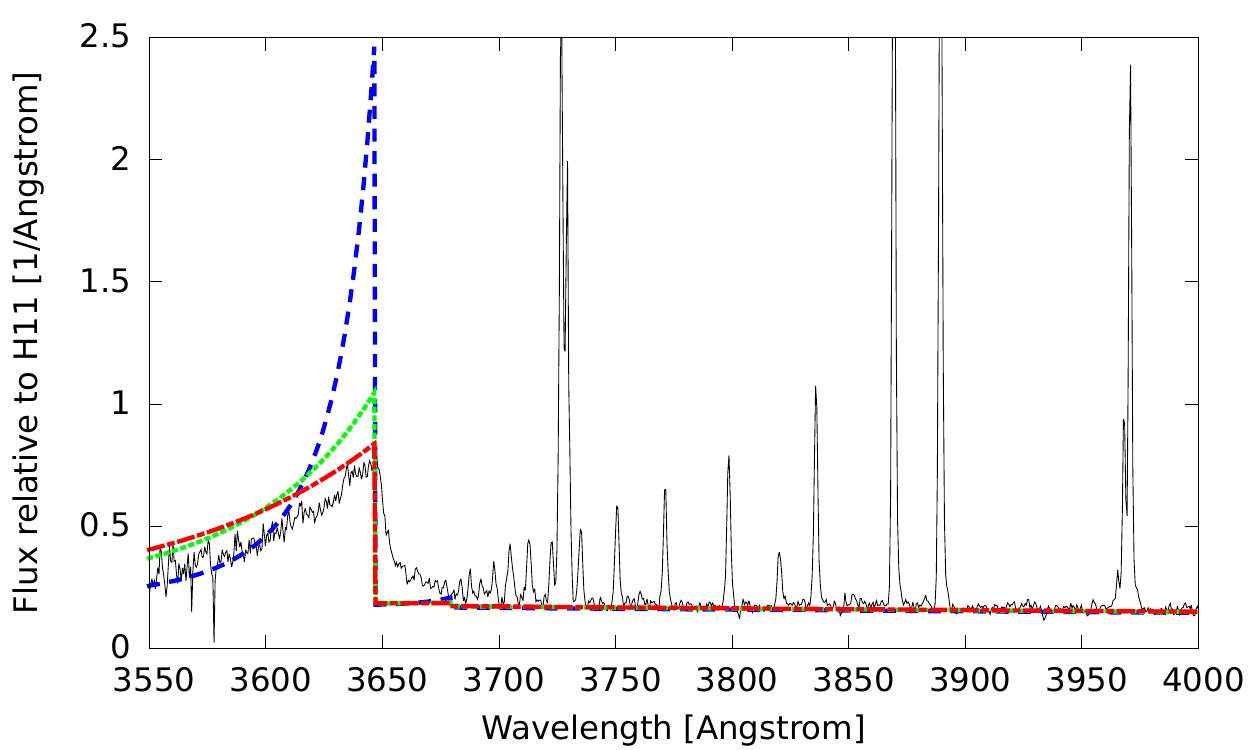}  \caption{Model continua with a $\kappa$-distribution of electron
energies for a temperature $T=1000$~K; $\kappa=2$ (dashed blue line), $\kappa=5$ (dotted green
line), and $\kappa=25$ (dot-dashed red line).} \label{T10003kappa}
\end{figure}

An alternative possible resolution of the ORL/CEL problem is that there is no separation of the
emitting material into high and low temperature components with significantly different abundances
but rather a single medium in which the electron energy distribution is non-Maxwellian, with a
$\kappa$-distribution being a possible candidate. The Balmer continuum close to the Balmer
discontinuity provides a means of testing the validity of this proposal for the very low energy
part of the distribution function. In Figures~\ref{T100003kappa} and \ref{T10003kappa} we
illustrate how the magnitude of the Balmer discontinuity and shape of the Balmer continuum change
as $\kappa$ is varied for two different temperatures. Compared to a MB, $\kappa$-distributions have
more particles at the lowest and highest energies and less at intermediate energies. At low
energies the $\kappa$-distribution function thus falls more rapidly with increasing energy than a
MB distribution. This can be seen in Figure~\ref{T10003kappa}, for example, with the magnitude of
the Balmer discontinuity increasing and the slope of the Balmer continuum becoming steeper as
$\kappa$ decreases.

\begin{figure}
\centering{}
\includegraphics
[scale=0.7] {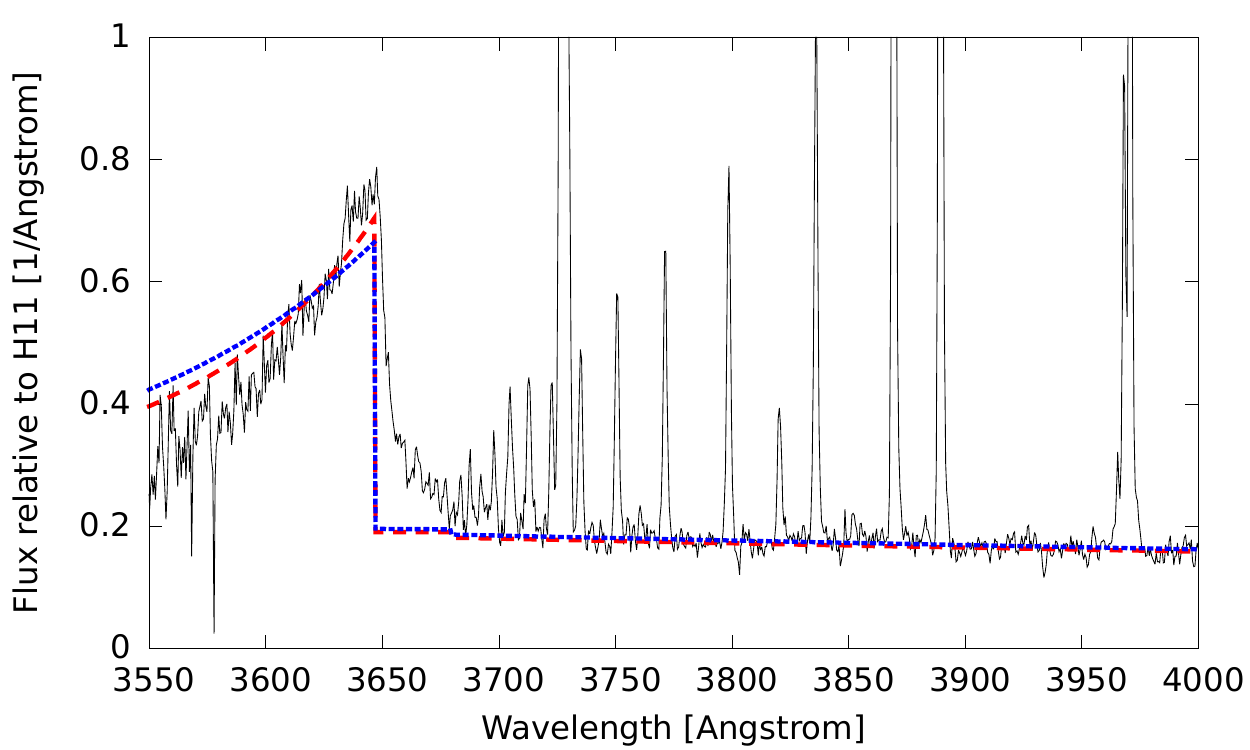} \caption{Fits obtained by simultaneous optimization of $\kappa$ and $T$
(dashed red line) and by fixing $\kappa=10$ and allowing $T$ to vary (dotted blue line).}
\label{1kappa}
\end{figure}

We now fit the chosen three spectral segments to a $\kappa$-distribution of which the temperature
and value of $\kappa$ are free parameters and the final values that we obtained for these
parameters are $\kappa=2.11$ and $T=4640$~K. Figure~\ref{1kappa} shows a plot of this fit. On
visual inspection the best-fit $\kappa$ distribution where $\kappa$ and temperature are allowed to
vary is significantly less good than the two-component MB fit. It should be remarked that the
optimized $\kappa$ obtained from this fitting exercise is very different to values proposed by
\citet{NichollsDS2012} which usually fall in the range 10-20. For comparison, we include in
Figure~\ref{1kappa} a fit carried out with a fixed value of $\kappa=10$. The temperature was
allowed to vary and the best fit value was $1665$~K. In the next section we attempt to quantify the
relative quality of these fits.

\subsection{Statistical Analysis} \label{Discussion}

We wish to address the question of whether the fit to the continuum using a $\kappa$-distribution
is significantly worse than the two-component MB fit. A visual inspection of Figures~\ref{2MB1free}
and \ref{1kappa} shows that while the-two-component fit provides a very good match to the three
segments of the continuum used for fitting, the best $\kappa$-distribution fit underestimates the
magnitude of the Balmer discontinuity and also underestimates the steepness of the decline at
shorter wavelengths. We will use a comparison of $\chi^2$ values for these two fits to quantify the
relative goodness of fit. To estimate $\chi^2$ we need an estimate of the uncertainties on the
observational data, which by inspection and from instrumental considerations are not independent of
wavelength, being much larger at the shortest wavelengths. We therefore compute separate rms
deviations of the data from the two-component MB fit for each of the three wavelength segments.
Since we use deviations from the two-component fit to define the data uncertainties it follows that
the value of $\chi^2$ for the fit, summed over the three wavelength segments,  will be given by
\begin{equation}
\chi^2 = N_{df} \equiv N_{dp} - N_{fp} = 237,
\end{equation}
where $N_{df}$  is the number of degrees of freedom, $N_{dp}$ is the number of data points in the
three segments and $N_{fp}$ is the number of free parameters. We now use the estimates of the data
uncertainties from the two-component MB fit to compute $\chi^2$ for the best fit
$\kappa$-distribution and we find $\chi^2=433$. We wish to test the null hypothesis that the
$\kappa$-distribution is an equally good fit to the data as the two-component MB fit. The following
square-root transformation of $\chi^2$ \citep{AbSteg72},

\begin{equation}
t = \sqrt{2\chi^2} - \sqrt{2N_{df}},
\end{equation}
is such that for large $N_{df}$, the transformed distribution is approximately normal with zero
mean and unit standard deviation. For the best fit $\kappa$-distribution we find that $t=7.67$
standard deviations from the zero mean, indicating that the probability that the null hypothesis is
true is vanishingly small.

Thus, a two-component \MB\ model is highly favored compared to the single-component $\kappa$ model.
The single-component MB model is even less likely to be correct than either of the two-parameter
models based on the same $\chi^2$ analysis, giving $t=14.4$ standard deviations.

\section{Conclusions} \label{Conclusions}

The analysis of the Balmer continuum spectrum of \hf\ seems to indicate the signature of a
two-component nebula with two different temperatures. Assuming that one component has a temperature
of 8800~K, we find that the best fit to the data is obtained by adding a second component with
temperature of 540~K and emission measure 15.0\% of the total. We also modeled the continuum with a
single component with $\kappa$-distributed free electron energies, but on the basis of a $\chi^2$
test we found this model to be significantly less likely to be correct. In the case of
$\kappa$-distribution models, it is important to note that modeling the Balmer continuum only
samples the free electron energy distribution at the lowest energies, below the mean energy, and
therefore gives no information about departures from MB at energies above the mean energy which
might affect the excitation of nebular forbidden lines, for example. The extreme nature of \hf\ and
the similarity of some of its spectral features to old novae like DQ~Her may also cast some doubt
on the applicability of the conclusions reached to the properties of planetary nebulae in general.

\section{Acknowledgment and Statement}

The work of PJS was supported in part by STFC (grant ST/J000892/1). During the final stages of
writing this paper, our attention was drawn to a similar paper by Y. Zhang, X-W. Liu and B. Zhang
(arXiv:1311.4974); however this did not have an impact on the methods or results reported in the
present paper.

\end{document}